\documentclass[notitlepage,twocolumn,superscriptaddress,amsmath,amssymb,aps,pra]{revtex4-2}
\usepackage{amsmath,amssymb,graphicx,mathtools,dsfont,ulem,bbold}
\usepackage[dvipsnames]{xcolor}
\usepackage[T1]{fontenc}
\usepackage{hyperref}
\usepackage{soul}
\usepackage{bibunits}

\usepackage{ulem}

\hypersetup{citecolor=red,colorlinks=true,urlcolor=blue}

\newcommand{\ket}[1]    {|#1 \rangle}
\newcommand{\ketbra}[2]{|#1\rangle\!\langle#2|}

\newcommand{\tr}[1]    {{\rm Tr}\left[ #1 \right]}
\newcommand{\av}[1]    {\langle #1 \rangle}

\newcommand{\modsq}[1]    {\left| #1 \right|^2}

\newcommand{\En}{\mathcal E_L}
\newcommand{\Enq}{\mathcal E_L^{(q)}}
\newcommand{\Q}{\mathcal Q_{L}^{(q)}}

\newcommand{\Enqtt}{\tilde{\mathcal E}_L^{(q)}(T)}
\newcommand{\Enqt}{\tilde{\mathcal E}_L^{(q)}}
\newcommand{\Qt}{\tilde{\mathcal Q}_{L}^{(q)}}

\newcommand{\Qm}{\mathcal Q_{L,{\rm max}}^{(q)}}

\begin{document}

\title{Inherent quantum resources in stationary spin chains}
 
\author{Marcin Płodzień} 
\affiliation{ICFO-Institut de Ciencies Fotoniques, The Barcelona Institute of Science and Technology, 08860 Castelldefels (Barcelona), Spain}

\author{Jan Chwedeńczuk} 
\affiliation{Faculty of Physics, University of Warsaw, ul. Pasteura 5, 02-093 Warszawa, Poland}

\author{Maciej Lewenstein}
\affiliation{ICFO-Institut de Ciencies Fotoniques, The Barcelona Institute of Science and Technology, 08860 Castelldefels (Barcelona), Spain}
\affiliation{ICREA, Passeig Lluis Companys 23, 08010 Barcelona, Spain}

\begin{abstract}
The standard way to generate many-body quantum correlations is via a dynamical protocol: an initial product state is transformed by interactions that generate non-classical correlations at later times. Here, we show that many-body Bell correlations are inherently present in the eigenstates of a variety of spin-1/2 chains. In particular, we show that the eigenstates and thermal states of the collective Lipkin-Meshkov-Glick model possess many-body Bell correlations. We demonstrate that the Bell correlations present in the eigenstates of the Lipkin-Meshkov-Glick model can take on quantized values that change discontinuously with variations in the total magnetization. Finally, we show that these many-body Bell correlations persist even in the presence of both diagonal and off-diagonal disorder.
\end{abstract}

\maketitle

\section{Introduction}

The system is said to be \textit{quantum} if any of its properties cannot be explained within the framework of classical physics. In complex systems these quantum deviations from classicality are
often manifested through  many-body correlations. The discussions among prominent figures such as Einstein~\cite{epr} and Schr\"odinger~\cite{sch} gave rise to concepts such as entanglement~\cite{horodecki2009entanglement, srivastava2024introduction,Horodecki2024}, Einstein-Podolsky-Rosen steering~\cite{uola2020steering} and, many years later, the correlations that are responsible for the violation of the postulates of local realism by some quantum systems~\cite{bell,brunner2014bell}. Modern quantum technologies, such as quantum computing, communication, or sensing~\cite{Frerot_2023}
are powered by these correlations and therefore depend on our control of these resources at all stages: namely the ability to generate, 
store and verify quantum correlations~\cite{PhysRevLett.117.210504, Acin_2018,Eisert2020,Kinos2021,PhysRevLett.125.150503,Laucht_2021,Zwiller2022,Fraxanet2022, Sotnikov2022, huang2024certifying,PRXQuantum.2.010201}.  

One of the main ways to generate strongly correlated quantum states is via dynamical protocols. Starting from a state in which all components of a quantum system are uncorrelated, 
their mutual interactions generate quantum correlations at later times. For example,
the analogue quantum simulators use the One-Axis Twisting (OAT) protocol~\cite{kitagawa1993squeezed,wineland1994squeezed}
to create strong and scalable quantum resources. The power of the OAT lies in its ability to generate spin-squeezed, many-body entangled, 
and many-body Bell-correlated states. 
The problem of generating nonlocality for the arbitrary depth $k$ has been addressed in~\cite{PhysRevLett.95.120502,Tura2014,Tura2014a,Tura2015,PhysRevA.100.032307,Baccarri2019,Aloy_2021,PRXQuantum.2.030329,PhysRevA.105.L060201} using a set of Bell inequalities based on second-order correlators, violation of which implies $k$-producibility of
nonlocality with $k\le6$ for large number of parties. Recently, Bell inequalities utilizing many-body correlators 
\cite{PhysRevLett.65.1838,PhysRevLett.100.200407,PhysRevLett.88.210401,PhysRevLett.99.210405, Reid2012, PhysRevA.84.032115,PhysRevLett.131.070201} were used to demonstrate that the OAT can actually generate arbitrary depth of nonlocality~\cite{plodzien2020producing,plodzien2022one,yanes2022one, PhysRevA.107.013311, PhysRevB.108.104301, plodzien2024generation,yanes2024exploring}.
The OAT protocol has been  studied theoretically and realised experimentally, 
and its relevance to quantum information science and high precision metrology has been demonstrated~\cite{doi:10.1126/science.aad8665,PhysRevLett.118.140401,pezze2018quantum,wolfgramm2010squeezed,muller2023certifying}.

In this work we focus on the stationary approach. Instead of analyzing many-body Bell correlations generated with dynamical  protocols, such as  the OAT discussed above, we show that many-body Bell correlations
are inherently available in the   eigenstates, and thermal equilibrium Gibbs states of spin-$1/2$ chains described by the family of the 
Lipking-Meshkov-Glick (LMG) model~\cite{LIPKIN1965188, MESHKOV1965199, GLICK1965211, PhysRevLett.49.478, LermaH_2014, PhysRevB.28.3955, Carrasco_2016, PhysRevA.97.012115}. 
We provide extensive analytical results supporting the numerical approach and allow the determination of the critical temperature above which the Bell correlations vanish.

The LMG model has attracted interest in recent years providing a versatile platform for proposals to realise 
quantum thermal machines and quantum heat engines~\cite{ PhysRevResearch.6.013310,PhysRevE.109.034112, HE20126594, PhysRevE.96.022143, HUANG2018604, Lobejko2020thermodynamicsof, Cakmak2020, PhysRevResearch.2.023145, Chen2019, centamori2023spinchain, Chakour2021, PhysRevA.106.032410, Solfanelli_2023, PhysRevB.109.024310}
The bipartite entanglement of the LMG model has been studied extensively over the last 
two decades~\cite{PhysRevA.70.062304, PhysRevA.71.064101, PhysRevLett.101.025701, SenDe2012, PhysRevA.85.052112,  PhysRevB.101.054431, Hengstenberg2023, BAO2020412297, Du_2011,PhysRevA.109.052618}, 
and the model itself has been realised on many physical platforms, such as Bose-Einstein condensates~\cite{Chen:09}, superconducting circuits~\cite{Larson_2010}, 
atoms in optical cavities~\cite{Sauerwein2023, 10.1063/1.5121558,PhysRevLett.100.040403,Grimsmo_2013,PhysRevA.97.023807}, or with quantum circuits~\cite{Hobday2023, hobday2024variance}.
The present work contributes to this rapidly developing field and supports the importance of the LMG systems for modern quantum technologies.
We note that the multipartite entanglement in a ground state of the Dicke model has recently been analyzed in~\cite{lourenco2024genuine}.

The paper is organised as follows. In the Sec.\ref{sec:Preliminaries} we provide preliminary overview of the work presented. In
Sec.\ref{subsec:Model}, we give an introduction to the considered LMG model, and in 
Sec.\ref{subsec:BellCorrelations}, we give an introduction to the many-body Bell correlator which characterizes the strength of Bell correlations of an arbitrary state of a spin-$1/2$ chain. Next, in Sec.\ref{sec:Results} we present results for three limiting cases of the LMG model with all-to-all couplings between the spin, and we present results for finite-range interactions. We conclude in Sec.\ref{sec:Conclusions}.

\section{Preliminaries}\label{sec:Preliminaries}

\subsection{The Lipkin-Meskhov-Glick model}\label{subsec:Model}

The LMG model describes a family of 1D chains of $L$ spin-$1/2$ particles depicted by the Hamiltonian
\begin{align}\label{eq:H_GLM}
    \hat H_\gamma = -\frac{1}{L}\sum_{i<j=1}^L\left(\hat{\sigma}_x^{(i)}  \hat{\sigma}_x^{(j)}  + \gamma \hat{\sigma}_y^{(i)} \hat{\sigma}_y^{(j)}\right) - h\sum_{i=1}^L\hat{\sigma}_z^{(i)}.
\end{align}
Here $\hat{\sigma}_n^{(i)}$ is the $n$th Pauli matrix of the $i$th spin ($n=x,y,z$, $i\in\{1\ldots L\}$), while the anisotropy parameter $\gamma$ and the magnetic field coupling constant $h$ differentiate 
the members of this family. The pair-wise interactions are of infinite range, which allows to express the Hamiltonian using the collective-spin operators
\begin{align}\label{eq.coll}
  \hat S_n=\frac12\sum_{i=1}^L\hat{\sigma}_n^{(i)}
\end{align}
as follows (up to an additive $L$-dependent constant)
\begin{align}\label{eq:H_GLM_sym}
    \hat H_\gamma = -\frac2L\left(\hat S_x^2+\gamma \hat S_y^2\right) - 2h\hat S_z.
\end{align}
The spectrum of this Hamiltonian is spanned by $L+1$ symmetric Dicke states, namely the eigenstates of the total spin operator $\mathbf{\hat{S}}^2=\hat S_x^2+\hat S_y^2+\hat S_z^2$
and, for instance, the $\hat S_z$, i.e.,
\begin{subequations}
  \begin{align}
    \mathbf{\hat{S}}^2|S,m\rangle & = S(S+1)|S,m\rangle\\
    \hat{S}_z\ket{S,m}&= m|S,m\rangle,
  \end{align}
\end{subequations}
where $S =L/2$ and $m\in\{-S,\ldots,S\}$.

Here we focus on the three emblematic cases of the LMG model, when the anisotropy parameter takes the values $\gamma=-1$, $\gamma=0$ and $\gamma=1$, giving the Hamiltonians 
(again, up to an additive constant)
\begin{subequations}
  \begin{align}
    &\hat H_{-1} = -\frac2{L}\left(\hat S_x^2-\hat S_y^2\right) - 2h\hat S_z,\\
    &\hat H_{0} = -\frac2L\hat S_x^2-2h\hat S_z,\\
    &\hat H_{+1}=\frac2L\hat S_z^2-2h\hat S_z.
  \end{align}
\end{subequations}

Particularly interesting is the $\gamma=1$,  where the spectrum is spanned by the Dicke states $\ket{S,m}$'s, giving the parabolic dependence of the eigenenergies on $m$, i.e., 
\begin{align}\label{eq.g1}
  E_m(h) & = \frac2Lm^2 - 2hm.
\end{align}
Its minimal energy is reached when $m$ is equal to
\begin{align}\label{eq:m_vs_h}
  m=m_0(h)\equiv I\left[\frac{hL}2\right],
\end{align} 
i.e., the closest integer (for even $L$) or half-integer ($L$ odd) to the minimum of this function.
The ground state is hence
\begin{align}
  \ket{\psi^{(h)}_0}=\ket{S, m_0(h)}
\end{align}    
For instance, when $h=0$, the ground state is
\begin{align}
  \ket{\psi^{(h=0)}_0}=\ket{S, 0}
\end{align}    
and the double-degenerated excited states lie symmetrically at $m$ positive and negative.
On the other hand,  when $h$ is so large that $m_0\geqslant S$ the ground-state is
\begin{align}\label{eq.large.h}
  \ket{\psi^{(h)}_0}=\ket{S, S}
\end{align}    
and the excited part of the spectrum is spanned by all the $m$'s smaller than $S$---an observation that will be relevant in the Results section, Sec.\ref{sec:Results}.

This concludes the preliminary section on the LMG model. We now introduce a correlation function that turns out to be a useful tool that allows us to certify the presence and the strength of many-body 
Bell correlations in this system.

\subsection{Bell correlations}\label{subsec:BellCorrelations}

When assessing the presence and the strength of Bell correlations in an $L$-body system, we must refer to a corresponding model that is consistent with the
postulates of local realism. Here, we assume that each of $L$ parties can measure a pair of observables $\sigma^{(k)}_x$ and $\sigma^{(k)}_y$,  each yielding binary outcomes, i.e., $\sigma_{x/y}=\pm1$,
with $k\in\{1\ldots L\}$. The correlation function (also referred to as the correlator) of these results is here defined as
\begin{align}\label{eq.el}
  \En=\modsq{\av{\prod_{k=1}^L\sigma_+^{(k)}}},
\end{align}
with $\sigma_+^{(k)}=1/2(\sigma_x^{(k)}+i\sigma_y^{(k)})$. This correlation is consistent with the local and realistic theory if the average can be expressed in terms of an integral
over the random (``hidden'') variable $\lambda$ distributed with a probability density $p(\lambda)$ as follows
\begin{align}\label{eq.lhv}
  \av{\prod_{k=1}^L\sigma_+^{(k)}}=\int\!\!d\lambda\,p(\lambda)\prod_{k=1}^L\sigma_+^{(k)}(\lambda)
\end{align}
We substitute this expression into Eq.\eqref{eq.el} and use the Cauchy-Schwarz inequality for complex integrals together with the fact that $\modsq{\sigma_+^{(k)}(\lambda)}=1/2$ for all $k$,  
giving
\begin{align}
  \En&=\modsq{\int\!\!d\lambda\,p(\lambda)\prod_{k=1}^L\sigma_+^{(k)}(\lambda)}\nonumber\\
  &\leqslant\int\!\!d\lambda\,p(\lambda)\prod_{k=1}^L\modsq{\sigma_+^{(k)}(\lambda)}=2^{-L}.
\end{align}
We thus conclude that
\begin{align}\label{eq.bell}
  \En\leqslant2^{-L}
\end{align}
is the $L$-body Bell inequality, because its violation defies the postulates of local realism expressed in Eq.~\eqref{eq.lhv}. It is valid for systems that yield binary outcomes of local, 
single-particle observables. This is the case of the $L$-spin LMG model, hence the quantum-mechanical equivalent of Eq.~\eqref{eq.el}, labeled with  ($q$) and defined as
\begin{align}\label{eq.elq}
  \Enq=\modsq{\av{\bigotimes_{k=1}^L\hat\sigma_+^{(k)}}},
\end{align}
will be used in the remainder of this work to witness Bell correlations. For convenience, we introduce
\begin{align}\label{eq.q.def}
  \Q=\log_2\left(2^L\Enq\right)
\end{align}
with which the Bell inequality from Eq.~\eqref{eq.bell} takes a more ``user-friendly'' form
\begin{align}\label{eq.bell.q}
  \Q\leqslant0.
\end{align}

The value of the Bell correlator ${\cal Q}^{(q)}_L$ provides information about the structure of the many-body quantum state and the strength of many-body Bell correlations. Namely, the maximal value of the correlator that is reachable by quantum mechanical systems is $\Qm=L-2$, which is obtained with the Greenberg-Horne-Zeilinger (GHZ) state
\begin{align}\label{eq.ghz}
    \ket{\psi_{\rm ghz}}=\frac1{\sqrt2}\left(\ket1^{\otimes L}+\ket0^{\otimes L}\right).
\end{align}
Next, all the values of $\Q$ that lie in the region $\Q\in]0,L-2]$ signal the presence of Bell correlations of different strength. In particular, when
\begin{align}\label{eq.depth}
  \Qm-(n+1)<\Q\leqslant\Qm-n
\end{align}
with $n\in\mathds N$, then \{the state is maximally $(n+1)$-local. This means that the state can be expressed in terms of maximally $n+1$ subsystems, mutually locally correlated. 
For instance, when the state is 2-local, the maximal value of the correlator is reached when $L-1$ parties form a quantum-mechanical GHZ state, giving the value of the correlator $\Enq=\frac14$
while the remaining one is
a particle measurements on which yield binary outcomes, just as duscussed below Eq.~\eqref{eq.el}. In this case, the correlator is 
$\Enq=\frac14\cdot\frac12$ giving
$\Q=\Qm-1$. 
Naturally, a 2-local state can have weaker correlations. For instance when the state is a product of two GHZ states of $L/2$ particles each, then $\Enq=\frac14\cdot\frac14$ giving $\Q=\Qm-2$. 
When the system is 3-local, then again the $\Q$ is maximal in the scenario, when $L-2$ particles form a GHZ state and the remaining two are mutually Bell-uncorrelated, giving $\Q=\Qm-2$.
Now the  pattern is established, i.e., a state characterized by maximally $(n+1)$-locality can reproduce Eq.~\eqref{eq.depth}. 
{  As a last example, we can consider a $L$-spins state being a product of $n$ separable parties, each yielding binary outcomes, and not restricted by quantum mechanics, and $k$ products of GHZ states, then $\Enq=\frac{1}{2^n}\cdot\frac{1}{4^k} = \frac{1}{2^{2k+n}}$.}
The concept of $k$-locality is a close analogy of $k$-separability of quantum states
~\cite{10.5555.2011464.2011472,Guhne_2010,PhysRevA.78.032101,Gao2011,Gao_2013,Szalay_2019,HONG2021127347}.
Note that the considered Bell correlator works for $L\geqslant3$ and for $n<L-2$. For $L=2$ the GHZ state from Eq.~\eqref{eq.ghz}, though Bell correlated, gives $\Q=0$. 
When $L$ is larger but $n=L-2$, then the state is $(L-1)$-local. Hence maximally a pair can form a Bell-correlated state and according to the above, this is not recognized by $\Q$, giving
according to Eq.~\eqref{eq.depth}: $\Q\leqslant0$. To witness two-body Bell correlations, other methods must be used~\cite{Tura2015}.

As a last remark of this introductory part, we stress that $\Enq$ (and hence $\Q$) is 
\begin{align}\label{eq.ghz.coh}
  \Enq=\modsq{\varrho_{\rm ghz}},
\end{align}
where $\varrho_{\rm ghz}$ governs the GHZ coherence between
all the spins $\ket1^{\otimes L}$ up and all down $\ket0^{\otimes L}$.
This observation---that information on Bell correlations is encoded in a single element of the density matrix---will be relevant in the forthcoming sections of this work.
This correlator belongs to a broader family considered in Ref.~\cite{zukowski2002bell}, where a general framework for certifying many-body Bell correlations was formulated. 
Its other variants include the Mermin form of Bell inequalities~\cite{mermin1990extreme,10.1119/1.12594},
the $N$-body inequalities~\cite{zukowski2002bell, cavalcanti2007bell, Reid2012, cavalcanti2011unified} or the Ardehali inequalities~\cite{Ardehali_1992}.
For a thorough discussion of Bell correlator considered here, we recommend 
the following works:~\cite{Plodzien2024EntanglementClassification,spiny.milosz,PhysRevLett.126.210506,PhysRevLett.129.250402,PlodzienGraph2024}.

\subsubsection{Bell correlator for symmetric states}

The LMG infinite-range model requires a symmetrized Bell correlator, i.e., such that uses the collective spin operators as defined in Eq.~\eqref{eq.coll}. The product of $L$ rising operators 
$\hat\sigma_+^{(k)}$ that address individual particles, as present in Eq.~\eqref{eq.elq}, should be replaced as follows
\begin{align}
  \bigotimes_{k=1}^L\hat\sigma_+^{(k)}\ \ \longrightarrow\ \ \hat S_+^L.
\end{align}
Note however, that when expressing the collective rising operator using Eq.~\eqref{eq.coll} we have
\begin{align}\label{eq.rising}
  \hat S_+^L=\left[ \sum_{k=1}^L\hat{\sigma}_+^{(k)}\right]^L,
\end{align}
which gives $L!$ times more ordered products of $L$ distinct operators. This combinatorial factor, stemming from the symmetry of the setup, must be taken into account.  
Hence the inequality~\eqref{eq.bell} adapted to the symmetric case (denoted by the tilde) reads
\begin{align}\label{eq:Epsilon_symmetric}
  \Enqt=\modsq{\frac1{L!}\av{\hat S_+^L}}\leqslant2^{-L}
\end{align}
or in analogy to Eq.~\eqref{eq.bell.q}
\begin{align}\label{eq.qt}
  \Qt=\log_2\left(2^L\Enqt\right)\leqslant0.
\end{align}
For a detailed discussion of the symmetrized many-body Bell correlators, see~\cite{10.21468/SciPostPhysCore.5.2.025}.

\subsubsection{Optimized correlator}

According to the previous sections, the correlator $\Q$ takes a product of $L$ operators rising the spin projection along the $z$-axis, each in the form of
\begin{align}
  \hat\sigma_+^{(k)}=\frac12\left(\hat\sigma_x^{(k)}+i\hat\sigma_y^{(k)}\right)
\end{align}
or, for the symmetrized case, the $\Qt$ is fed with
\begin{align}
  \hat S_+=\hat S_x+i\hat S_y.
\end{align}
Naturally, none of the arguments invoked above leading to the formulation of the Bell inequality~\eqref{eq.bell.q}, does change upon a choice of different orientations of local axes.
We can always prepare the rotation of the rising operator for the $k$-th spin,
\begin{equation}
    \hat{\sigma}_+^{(k)}  \to \hat{R}_k^\dagger(\vec{\theta}_k) \hat{\sigma}_+^{(k)} \hat{R}_k(\vec{\theta}_k),
\end{equation}
where $\hat{R}_k(\vec{\theta}_k) = e^{-i\vec{\theta}_k\cdot \vec{\sigma}_k}$,  $\vec{\theta}_k = \{\theta^{(k)}_x, \theta^{(k)}_y, \theta^{(k)}_z\}$, $\vec{\sigma}_k = \{\hat{\sigma}_x^{(k)}, \hat{\sigma}_y^{(k)}, \hat{\sigma}_z^{(k)} \}$, 
or for the symmetrized case
\begin{equation}\label{eq:S_theta}
    \hat{S}_+  \to \hat{P}^\dagger(\vec{\theta}) \hat{S}_+ \hat{P}(\vec{\theta}),
\end{equation}
$\hat{P}(\vec{\theta}) = e^{-i\vec{\theta}\cdot \vec{S}}$,  $\vec{\theta} = \{\theta_x, \theta_y, \theta_z\}$, $\vec{S} = \{\hat{S}_x, \hat{S}_y, \hat{S}_z \}$. 
This freedom of choice allows us to optimize the correlator $\Q$,
i.e., adapt it to the geometry of the many-body state that is under consideration.

In practice the correlator is optimized numerically and its maximal value is found. For the LMG model as considered here, the permutational symmetry of the Dicke states
implies that the optimization procedure is collective. However, in general, when such spin-exchange symmetry is not present, the axes must be
determined individually for each particle.

In the following section, we present the optimized correlator $\Qt$ for the LMG eigenstates.
Before we discuss the results, let us stress again that the many-body Bell correlations are encoded in the GHZ coherence term of the density matrix, see Eq.~\eqref{eq.ghz.coh}, and that 
the optimization of the correlator (Heisenberg picture) is equivalent to rotating the density matrix itself (Schr\"odinger picture).

\section{Results}\label{sec:Results}

With an appropriate tool for the study of spin-$1/2$ chains at hand, we can start to analyze the many-body Bell correlations in the eigenstates of LMG models.

\subsection{Pure states}\label{sec:Results_pure_states}

For a given set of parameters $\{L, \gamma, h\}$ we obtain the set of eigenvalues and eigenvectors of the Hamiltonian, Eq.\eqref{eq:H_GLM}, 
$\hat{H}_\gamma |\psi^{(h)}_v\rangle = E^{(h)}_v |\psi^{(h)}_v\rangle$, $v = 0\dots L$, and for each eigenvector $|\psi^{(h)}_v\rangle$ we find the optimized value of the correlator $\Enqt$, Eq.\eqref{eq:Epsilon_symmetric}, denoted as $\Enqt(v, h)$, and the corresponding $\Qt(v, h)$, Eq.\eqref{eq.qt}.

\subsubsection{Numerical analysis}

\begin{figure}
  \includegraphics[width=\linewidth]{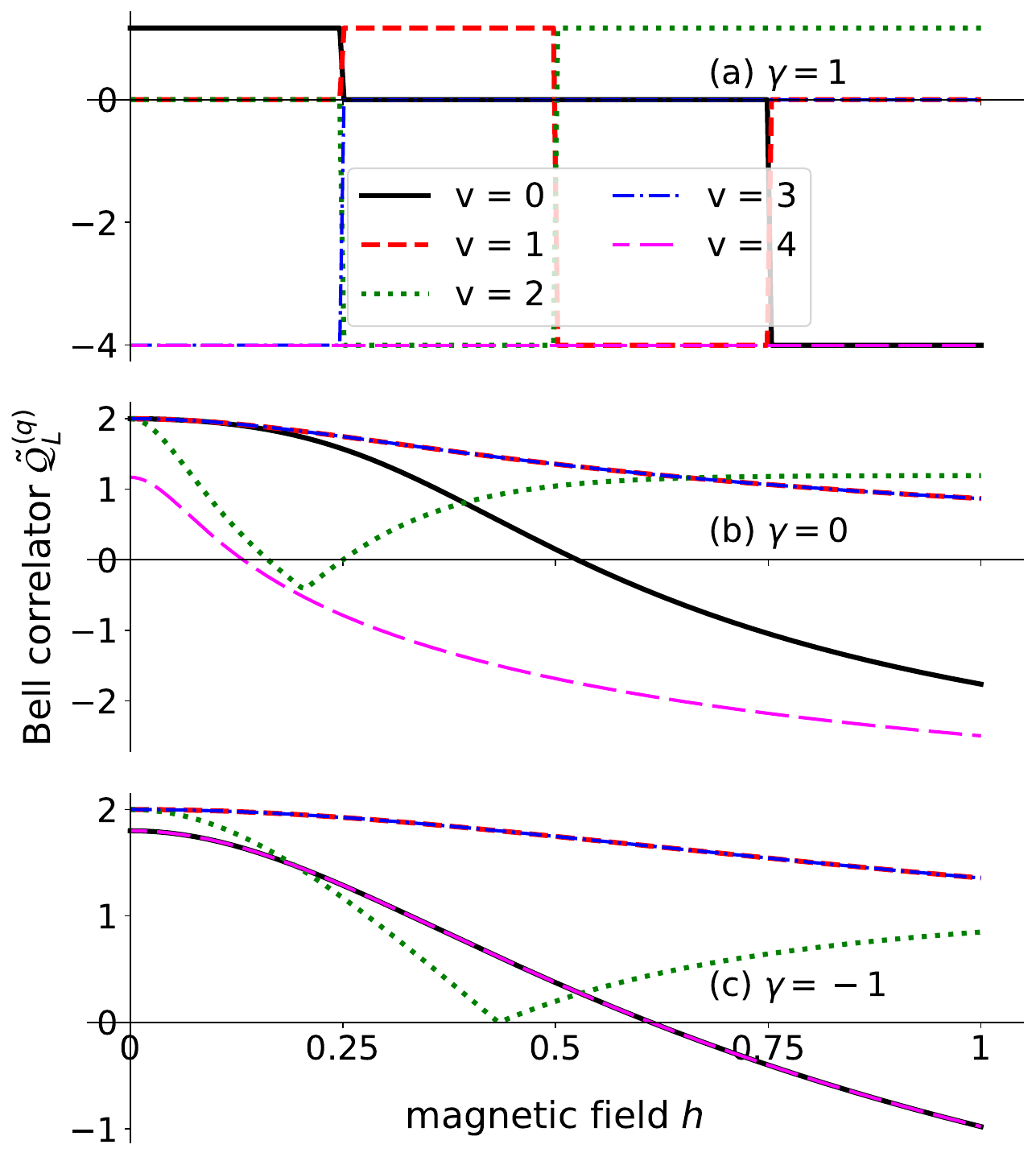}
  \caption{Value of the optimized Bell correlator, $\Qt$, Eq.\eqref{eq.qt} for an eigenstate $|\psi^{(h)}_v\rangle$ of LMG Hamiltonian $\hat{H}_{\gamma}$, as a function of the magnetic field for $L = 4$ spins. The black solid line corresponds to the ground state ($v=0$), the red dashed line corresponds to the first excited state ($v = 1$), the green dotted line to the $v=2$, the blue dash-dot line to $v=3$, and the magenta long-dashed line to $v=4$.   }
\label{fig:fig_1}
\end{figure}

We start our analysis with the simple case of $L = 4$ spins, $S = 2$. 
In Fig.\ref{fig:fig_1} we present optimized Bell correlator $\Qt$ for eigenstates $|\psi_v\rangle$ of the $\hat{H}_\gamma$, as a function of the magnetic field $h$, $\gamma = \{1, 0, -1\}$. 
For each $\gamma$ the corresponding eigenstates reveal many-body Bell correlations, $\Qt>0$, for particular values of the magnetic field $h$. 

Note that the cases of $\gamma = 0$ and $-1$, displayed in Fig.\ref{fig:fig_1}(b)-(c), are characterized with continuous and smooth variations of Bell correlators calculated with an eigenstate $|\psi^{(h)}_v\rangle$. However, when $\gamma = 1$, so that $\hat{H}_{\gamma = 1} \propto \hat{S}_z^2$ , the many-body Bell correlations, as a function of $h$ for a given $|\psi_v^{(h)}\rangle$ are non-continuous, see Fig.\ref{fig:fig_1}(a). 

To explain the origin of these jumps 
let us focus the ground state $|\psi^{(h)}_0\rangle$.  
When $h = 0$ it takes the form $|\psi^{(h)}_0\rangle = |S, 0\rangle$, with magnetization $m = 0$. The growing $h$ does not change the ground state of the system as long as $h<1/4 = h_1 $. When $h \ge h_1$, it suddenly changes to $|\psi^{(h_1)}_0\rangle = |S, 1\rangle$, $m = 1$, and the optimized Bell correlation  jumps to $\Qt(0, h_1)$. Again, the ground state remains unchanged as long as $h< h_2 = 3/4$ and once this value is passed we get $|\psi^{(h_2)}_0\rangle = |S, S\rangle$, $m = 2$, and Bell correlations jump to the value $\Qt(0, h_2)$.

As we can observe, a change in $h$ implies the jump of the magnetization  in the ground state, which determines the value of the Bell correlations $\Qt(0, h_2)$. Equivalently, we can say that for every ground state $|\psi^{(h)}_0\rangle$ there exists an $v$-th eigenstate $|\psi^{(h=0)}_v\rangle$ corresponding to $h=0$, i.e. for each $h$ we can always can find index $v$ such that  $|\psi^{(h)}_0\rangle = |\psi^{(h=0)}_v\rangle$, thus $\Qt(0, h) = \Qt(v, 0)$. 
These simple observations allow for fully analytical results, as we now show.

\subsubsection{Analytical expression for $\Enqt$ for $\gamma = 1$}

{

The one-axis twisting protocol (governed by the LGM model with $\gamma = 1$) dynamically generates the many-body entanglement and Bell correlations. Starting from the initial state product state $|\psi_{\rm ini}\rangle=e^{-i\frac{\pi}{2}\hat{S}_y}|S,S\rangle$, the time evolved state
$|\psi(t)\rangle = \sum_{m=-L/2}^{L/2}e^{-i t E_m(h=0)}\langle\psi_{\rm ini}|S,m\rangle|S,m\rangle$ is entangled, where $E_m(h=0)=2m^2/L$, Eq.\eqref{eq.g1}. At a specific time $t_{\rm GHZ}$ a GHZ-state in the $x$-direction is generated \cite{plodzien2020producing,plodzien2022one}. As such, the correlator $\Enqt$ is maximized when the spin rising operators are oriented along the $x$-direction, and the optimal orientations 
 $\vec{\theta}^* = \arg\max_{\vec{\theta}}\Enqt(\vec{\theta})
$ are $\vec{\theta}^* =\{0,0,0\}$
\footnote{Formally solving $
    \frac{\partial}{\partial\vec\theta}\Enqt(\vec{\theta}) = 0$,
for $\vec{\theta}$ we get $\vec\theta = \{0, 0, 0\}$,  for each eigenstate of $\hat{H}_{\gamma=1}$.}.
This fact, allows us to anticipate that for all eigenstates of the $\hat{H}_{\gamma = +1}$ Hamiltonian, the optimal orientation of the two axes defining the rising operator are
\begin{align}\label{eq.splus}
  \hat S^{(x)}_+=\hat S_{y}+i\hat S_z,
\end{align}
which we also confirmed numerically ~\footnote{
From now on, to distinguish between the different orientations of the rising operators, we add the proper superscript [here $(x)$]. }.
}


It is now our goal to express the expectation value of the Bell correlator fed with operator from Eq.~\eqref{eq.splus} on the Dicke states. These are the eigenstates of $\hat S_z$ hence the calculation requires an additional rotation through the $y$-axis that would align the $x$ and $z$ orientations, i.e., 
\begin{align}\label{eq.splus.rot}
  \hat S^{(x)}_+=e^{-i\frac\pi2\hat S_y}\hat S^{(z)}_+e^{i\frac\pi2\hat S_y}.
\end{align}
The action of such a rotation operator on the Dicke state is
\begin{align}\label{eq.wign}
  e^{i\frac\pi2\hat S_y}\ket{S,m}=\sum_{m'=-S}^Sd_{m;m'}^{S}\left(\frac\pi2\right)\ket{S,m'},
\end{align}
where $d_{m;m'}^{S}$ represents an element of the Wigner matrix in the fixed-$S$ subspace~\footnote{See~\cite{varshalovich1988quantum} for the details on the rotation matrices in the SU(2) group}. 
As argued below Eq.~\eqref{eq.ghz.coh}, the Bell correlator is fully determined by the GHZ-coherence, hence by the modulus square of the product of the coefficients multiplying the
extreme elements $m'=S$ and $m'=-S$ in Eq.~\eqref{eq.wign}, namely
\begin{align}\label{eq:Epsilon_analytical_binomial}
  \Enqt=\modsq{d_{m;S}^{S}\left(\frac\pi2\right)d_{m;-S}^{S}\left(\frac\pi2\right)}=\left[\frac{1}{2^{L}} \binom{L}{m + \frac{L}{2}}\right]^2.
\end{align}
When $L$ is large, {   and $|m|\ll L$}, the binomial can be approximated with the Gaussian, giving
\begin{equation}\label{eq:correlator_analytical}
    \Enqt \simeq\frac2{\pi L}e^{-4\frac{m^2}L}.
\end{equation}
For $m=0$, the Bell limit is surpassed for all $L>3$.
It is always maximal at $m=0$, hence it is not necessarily the ground state that is most non-classical. In particular, for sufficiently large $h$, the 
ground state given by Eq.~\eqref{eq.large.h} is separable---all the $L$ spins point in the upward direction. 

\begin{figure}[t!]
    \centering
    \includegraphics[width=\linewidth]{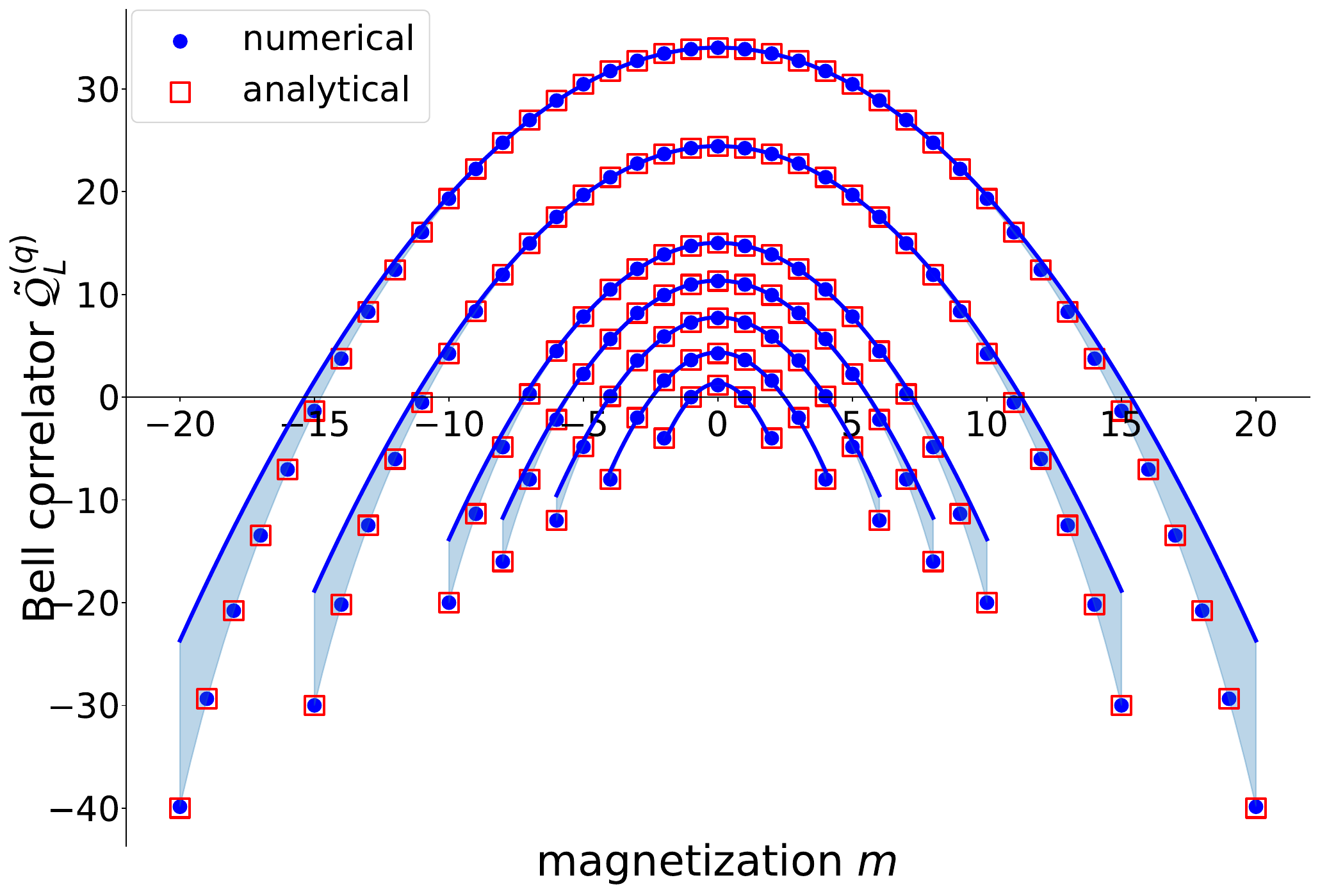}
    \caption{
    The optimized Bell correlator $\Qt$ (blue dots) for an eigenstate $|\psi^{(h)}_v\rangle$ of LMG Hamiltonian $\hat{H}_{\gamma}$, $\gamma = 1$, as a function of the magnetization $m = \langle \hat{S}_z\rangle$ for $L = 4, 8, 12, 16, 20, 30, 40$ (from inner to outer lines,  respectively).  Red squares represent the analytical {   exact} expression, Eq.\eqref{eq:Epsilon_analytical_binomial}, {   while solid blue lines correspond to continuous approximation, Eq.\eqref{eq:correlator_analytical}}. {   Shaded areas represent discrepancy between continious approximation and exact results for increasing $|m|$.}}
    \label{fig:Q_analytical}
\end{figure}

To verify our analytical findings, in Fig.~\ref{fig:Q_analytical}, we present 
$\Qt = \log_2(2^L\Enqt)$, Eq.\eqref{eq.qt}, as a function of magnetization $m = \langle \hat{S}_z\rangle$. For each set of parameters $\{ L, h\}$ we calculate all the eigenstates $|\psi_v\rangle$ of $\hat{H}_{\gamma = +1}$ and numerically confirm the optimal orientation of Eq.~\eqref{eq.splus}. We plot $\Qt$ as a function of $m$, with the blue dots presenting the numerical optimization of the correlator, while the red diamonds corresponding to our analytical findings, Eq.~\eqref{eq:Epsilon_analytical_binomial}. The inverted parabola, {   solid blue lines, correspond to continuous approximation to} $\Qt(m)$, {   with Eq.\eqref{eq:correlator_analytical}}. {   These results} show that maximal Bell correlations are for the eigenstates for which the magnetization is zero, one of important results of the presented work.
Note that the normalized Bell correlator $\Qt$ can be linked with the energy from Eq.~\eqref{eq.g1}, calculated at $h=0$, as follows
\begin{align} 
   \Qt=-\alpha E_m+\beta L+\gamma,
 \end{align}
 where $\alpha=2/\ln2$, $\beta=1$ and $\gamma=\log_2\left(\frac2{\pi L}\right)$. This is a simple formula but of high predictive power, showings the benefit of keeping
a string of spins at low energies/temperatures.
Note that the Bell correlations are present mostly in the zero- (or close to) magnetization sector; this observation is in an agreement with results of Ref.~\cite{Tura2015}, where Authors have identified classes of Bell inequalities for which zero magnetization states are optimal.
It is important to note that the Dicke states have been experimentally realized for few qubits using various physical platforms, including photonic systems~\cite{PhysRevLett.109.173604, PhysRevLett.103.020504, PhysRevA.83.013821, Zhao:15, Wang2016, Mu:20}, ultracold atoms~\cite{PhysRevLett.112.155304}, and quantum circuits~\cite{Chakraborty2014, 10.1007/978-3-030-25027-0_9, 9275336, PhysRevLett.103.020503, 9951196, Narisada2023}.

\subsection{Thermal states}
The next question we answer is about the impact of thermal fluctuations on the Bell correlations in the LMG models. 
To investigate this problem, we calculate the thermal density matrix
\begin{align}
  \hat{\varrho}_T = \frac{1}{{\cal Z}}\sum_{v} e^{-E_v/T}\ketbra{\psi_v}{\psi_v},
\end{align}
for each $\gamma$, where ${\cal Z} = \sum_v e^{-E_v/T}$ is the statistical sum and the summation runs over the whole spectrum. Next, we calculate the expectation value of the many-body Bell correlator, 
\begin{equation}\label{eq.Correlator_thermal}
  \Enqtt =  \big|\frac{1}{L!} \tr{\hat{\varrho}_T \hat{P}^\dagger(\vec{\theta})\hat{S}_+^{(x)L} \hat{P}(\vec{\theta})}\big|^2
\end{equation}
 optimized over $\vec{\theta}$.

\begin{figure}[t!]
    
    \includegraphics[width=\linewidth]{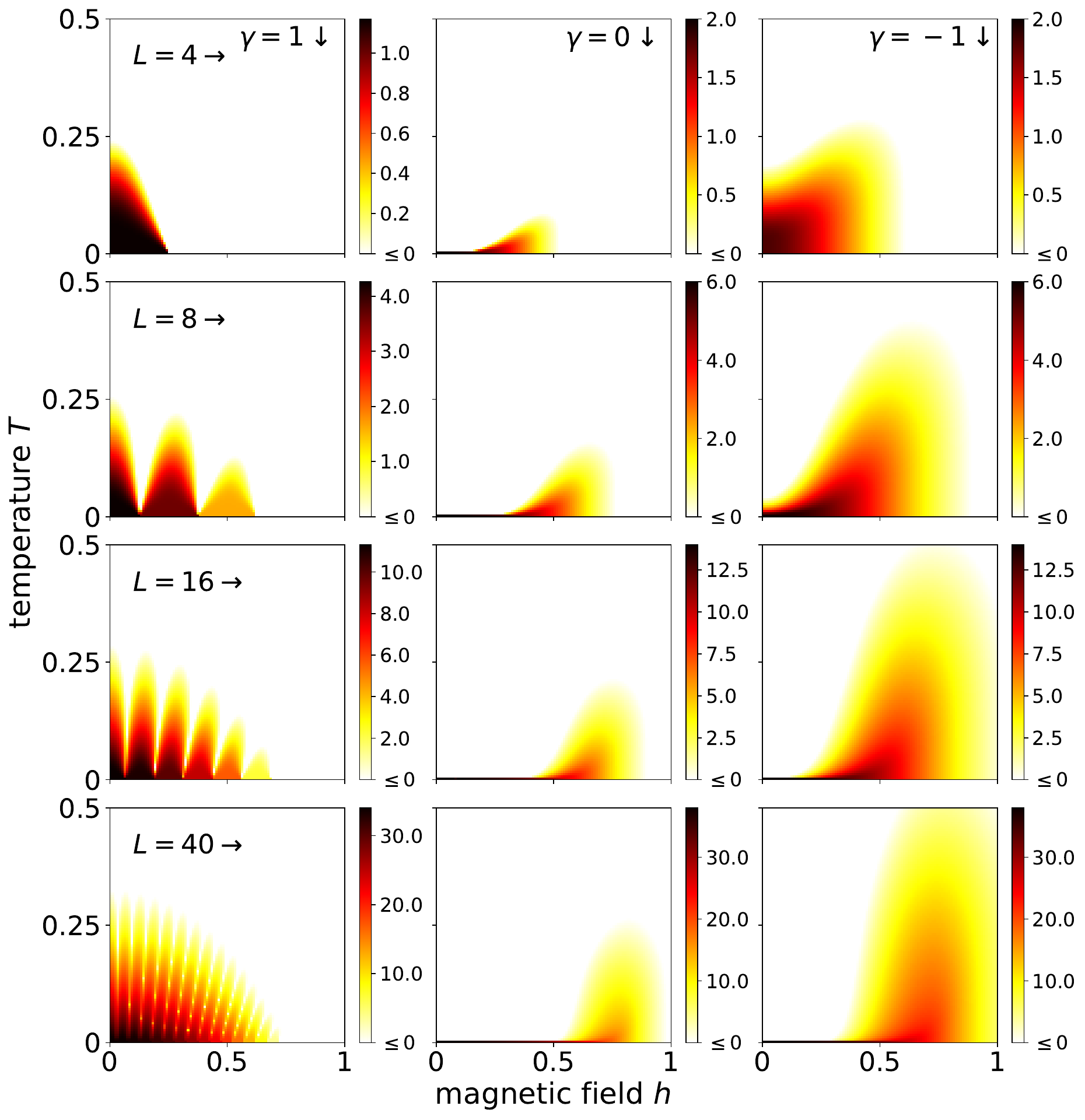}
    \caption{Color-encoded many-body Bell correlator $\tilde{{\cal Q}}^{(q)}_L$ for Gibbs state at temperature $T$ as a function of the magnetic field $h$ for $\gamma = \{1, 0, -1\}$ (from left to right, respectively) for $L = 4,8,16,40$. White background denotes non-positive values.}
    \label{fig:fig_3}
\end{figure}

In Fig.~\ref{fig:fig_3} we show the $\Qt$, Eq.~\eqref{eq.qt}, in the $h-T$ plane density plot for $\gamma = 1, 0, -1$ for given $L$. We observe that for a given temperature, the Bell correlator plotted as a function of $h$ shows a complicated and rich structure, knowledge of which could be relevant for future applications where the performance of a particular task relies on the strength of Bell correlations.

In the following paragraphs, we analytically investigate the origins of those fine structures for the $\gamma = 1$ and present qualitative analysis for $\gamma = 0,-1$. 

\begin{figure}[t!]
    \includegraphics[width=\linewidth]{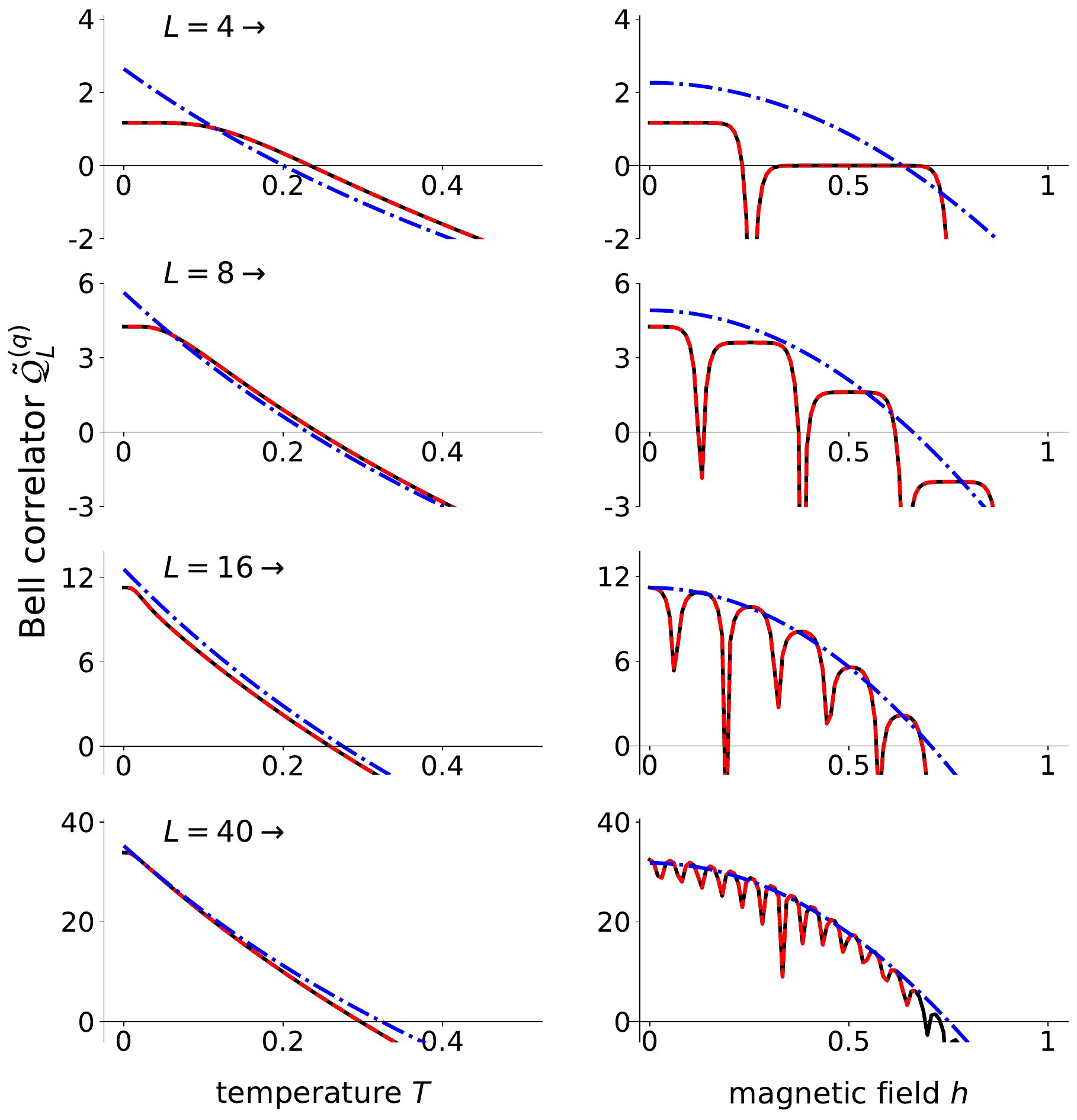}
    \caption{Many-body Bell correlator $\tilde{{\cal Q}}^{(q)}_L$ as a function of temperature $T$ (left column) for fixed magnetic field $h = 0.04$, and as a function of magnetic field $h$ for fixed temperature $T = 0.025$, $\gamma = 1$. Rows correspond to system size $L = 4, 8, 16, 40$ (from top to bottom, respectively). The black lines correspond to numerical results; red dashed line is the analytical formula from Eq.\eqref{eq:Epsilon_at_thermal_sum}, while the blue dashed-dotted line is the formula obtained in the $L\to\infty$ limit, Eq.\eqref{eq:Epsilon_envelope}.
    }
    \label{fig:fig_4}
\end{figure}

\subsubsection{Bell correlator for $T\geqslant0$ and $\gamma=1$ case}

For the $\gamma=1$ case, when the spectrum is spanned by the Dicke states, we have
\begin{equation}\label{eq.dicke.T}
    \hat{\varrho}_T= \frac{1}{{\cal Z}} \sum_{m=-L/2}^{L/2} e^{-E_m/T}|S,m\rangle\langle S,m|,
\end{equation}
where the energies are given by Eq.~\eqref{eq.g1}. Using Eq.~\eqref{eq:correlator_analytical}
we obtain 
\begin{equation}\label{eq:Epsilon_at_thermal_sum}
  \Enqtt \simeq \frac{2}{\pi L}\frac{1}{{\cal Z}^2}\big|\sum_{m=-L/2}^{L/2} (-1)^m e^{-\frac{2}{T}\left(\frac{m^2}L-hm\right)}e^{-\frac{2}{L}m^2}\big|^2,
\end{equation}
In Fig.\ref{fig:fig_4} we present the fixed-$T$ and fixed-$h$ cuts of the phase diagram shown in Fig.\ref{fig:fig_3}. 
The black solid lines corresponds to numerical results, while the red dashed lines to analytical expression, Eq.\eqref{eq:Epsilon_at_thermal_sum}, presenting a perfect agreement even for small $L$. Some more insight into the behavior of $\Enqtt$ from Eq.~\eqref{eq:Epsilon_at_thermal_sum}
can be retreived from the large-$L$ limit, where the sum over discrete $m$ is replaced with an integral. Since the $m$'s range from $-L/2$ to $L/2$, extending the corresponding
integral limits to $\pm\infty$ requires that the width of the $T$-dependent Gaussian $\sigma=\sqrt{LT}$ remains much smaller than $L$. 
Hence for the validity of the continuous approximation to hold, the following must be satisfied $T\ll L$.
In this regime, the Gaussian integral gives a straightforward outcome
\begin{align}\label{eq:Epsilon_envelope}
  \Enqtt\simeq{\frac{\pi}{2}}\frac{1}{ L(T+1)}e^{-\frac{h^2L}{1+T}}e^{-\frac{\pi^2LT}{4(1+T)}}.
\end{align}
In Fig.\ref{fig:fig_4}, we plot $\Qt$ for the above equation as a blue dash-dotted line, which gives an estimate for the envelope for $\Enqtt$, improving as $L$ grows. 
The structures seen in Figs~\ref{fig:fig_3} and~\ref{fig:fig_4} as fast oscillations of the correlator as a function of $h$ are lost in the approximation~\eqref{eq:Epsilon_envelope}. They
are a consequence of discrete-$L$ interference effects in Eq.~\eqref{eq:Epsilon_at_thermal_sum}. Clearly, as $L$ grows, these structures become denser and finer, and---as expected---are smoothed-out in
the $L\gg1$ limit.

The formula~\eqref{eq:Epsilon_envelope} allows for an estimate of the 
critical temperature $T_{\rm crit}$, above which the Bell correlations vanish for a given $h$.  We insert the expression~\eqref{eq:Epsilon_envelope} into the inequality Eq.~\eqref{eq.bell.q} and obtain that 
\begin{align}
  T_{\rm crit}=\frac{\ln2-h^2}{\frac{\pi^2}4-\ln2}
\end{align}
which is in a very good agreement with numerical results, visible in Fig.~\ref{fig:fig_4}. We stress that the critical temperature does not depend on the system size $L$, as expected from
the thermodynamic limit. In particular, when $h=0$ the critical temperature is maximal and equal to $T_{\rm crit}\simeq0.39$. Moreover, the condition $T_{\rm crit}=0$ gives the maximal $h=\sqrt{\ln2}\simeq0.83$
above which the Bell correlations vanish for all $T$, see the blue line in Fig.~\ref{fig:fig_4}. Note that the $T_{\rm crit}$ obtained from the many-body Bell correlator considered in this work is about $2$ times smaller than the critical temperature reported in~\cite{Fadel2018bellcorrelations}, based on Bell inequality formed with two-body correlators only.

\subsubsection{Bell correlator for $T\geqslant0$ and $\gamma=0$ case}

Some insight into the structure of the $T-h$ graph shown in Fig.~\ref{fig:fig_3} 
can be obtained also for $\gamma=0$. In this case the LMG system can be modelled with an approximate Schr\"odinger-like equation, as explained below.
We write the stationary Schrodinger equation, acting with the Hamiltonian $\hat H_{\gamma=0}$ 
on the state decomposed in the basis of eigen-states of $\hat S_x$ operator, namely
\begin{align}
  \ket\psi=\sum_{m=-L/2}^{L/2}\psi_m\ket{S,m}
\end{align}
which, projected onto $\ket{S,m}$, reads
 
  \begin{align}\label{eq.schr}
    h L\left[\psi_{m+1}f_+(z_m)+\psi_{m-1}f_-(z_m)\right]-\frac{L}{2}\psi_m z^{2}_{m}=E\psi_m.
  \end{align}

Here, we introduced a normalized variable 
\begin{align}\label{eq.def.z}
  z_m=2\frac{m}{L}
\end{align}
that changes between -1 and 1. Also, the coefficients $f_\pm(z_m)$ present in Eq.~\eqref{eq.schr} are defined as follows
 
  \begin{align}
    f_\pm(z_m)=\sqrt{\frac{1\mp z_m}{2}\left( \frac{1\pm z_m}{2}+\frac{1}{L}\right)}.
  \end{align}
 
For large $L$, the $1/L$ terms in these square-roots can be neglected and $z_m$ becomes quasi-continuous, allowing to approximate
the difference between the state coefficients $\psi_{m\pm1}$ with the second derivative over $z$. This gives a Schr\"odinger-like equation for the wave-function $\psi_m\rightarrow\psi(z)$ of a 
fictitious unit-mass particle
\begin{align}\label{eq.cont}
  \left(-\left(\frac{2}{L}\right)^2\sqrt{1-z^2}\frac{d^2}{dz^2} +V_{\rm eff}(z)\right)\psi(z)=\tilde E\psi(z),
\end{align}
where the effective potential is 
\begin{align}\label{eq.pot}
  V_{\rm eff}(z)=-\sqrt{1-z^2}+z^2\kappa/2
\end{align} 
and $\kappa=\frac1h$. The details of this derivation can be found in Ref.~\cite{ZP2008}. 

Crucially, the spectrum of the Eq.~\eqref{eq.cont} depends on the value of $\kappa$. When it is large, the second term of Eq.~\eqref{eq.pot} dominates giving an effective potential that is close
to harmonic. Interestingly, when $\kappa$ surpasses the threshold value $\kappa=-1$, the potential develops two minima located at $\pm z_0$, where $z_0=(1-1/\kappa^2)^{1/2}$ and becomes
\begin{align}\label{eq.pot.harm}
  V_{\rm eff}(z)\simeq\frac12\kappa(1-\kappa^2)(z\pm z_0)^2+\mathrm{const.}
\end{align}
The ground state tends to localize around these minima and form a macroscopic superposition
\begin{align}\label{eq.gr}
  \psi_{\rm gr}\simeq\frac1{\sqrt2}\left[\psi_0(z-z_0)+\psi_0(z+z_0)\right],
\end{align}
where $\psi_0(z\pm z_0)$ is the Gaussian ground state of each of the the harmonic contributions to Eq.~\eqref{eq.pot.harm}. 
When this superposition forms, the Bell correlator $\Qt$ becomes large, as it is exactly this type of macroscopic superposition that it is sensitive to, see Eq.~\eqref{eq.ghz.coh}. 
This is why the central column of Fig.~\ref{fig:fig_3} shows non-vansing Bell correlator when $h$ is sufficiently small.

However, an important feature of this plot is yet to be explained---namely why the $\Qt$ seems more vulnerable to thermal fluctuations when $h$ is very small (i.e., the area of non-zero Bell correlations
shrinks with $h\rightarrow0$).

To explain this behavior, note that as $h$ drops, so that $\kappa\rightarrow-\infty$,
the ground state~\eqref{eq.gr} becomes (assymptotically) degenerate with the first excited state 
\begin{align}\label{eq.1st}
  \psi_{\rm 1^{st}}\simeq\frac1{\sqrt2}\left[\psi_0(z-z_0)-\psi_0(z+z_0)\right].
\end{align}
In this regime, any non-zero value of $T$ will lead to occupation of both these states. The GHZ coherences of Eqs~\eqref{eq.gr} and~\eqref{eq.1st} have opposite signs, therefore
the Bell correlations quickly degrade in this thermal mixture. This is a qualitative explanation, for a more quantitative analysis, see~\cite{hamza2024bell}. 
When $h\simeq0$, the ground state is the GHZ state, giving the maximal Bell correlator, but only at $T=0$. This is represented
by a very thin and dark line in the central column of Fig.~\ref{fig:fig_3}. 

\subsubsection{Bell correlator for $T\geqslant0$ and $\gamma=-1$ case}

Finally, we provide the qualitative explanation for the phase diagram in the $\gamma = -1$ case. For this case, approximate analytical methods are not available. Nevertheless,
the exact diagonalization of the Hamiltonian $\hat H_{\gamma=-1}$ shows that as $L$ grows, its spectrum closely resembles that of $\hat H_{\gamma=0}$. In particular, for small $h$, the ground-
and first-excited states of $\hat H_{\gamma=-1}$ are almost degenerate and are close to those from Eq.~\eqref{eq.gr} and~\eqref{eq.1st}, respectively. This explains why the phase diagram 
for $\gamma=-1$ [right column of Fig.~\ref{fig:fig_3}] closely resembles that for $\gamma=0$ (central column). In particular, the structure of the two lowest-lying states
implies that for $\gamma=-1$ the Bell correlator is suspectible to non-zero temperature for small $h$.

\subsection{Distance-dependent interactions}
So far, we considered the LMG model, which is characterized by the constant strength all-to-all interactions, see Eq.~\eqref{eq:H_GLM}. We now take a step further and allow for the two-body coupling to depend
on the distance between the spins, by considering the following Hamiltonian
\begin{align}\label{eq:H_power_law}
  \hat{H} = -\frac{1}{L}\sum_{i<j}J_{ij}\left(\hat{\sigma}_x^{(i)}\hat{\sigma}_x^{(j)}+\gamma\hat{\sigma}_y^{(i)}\hat{\sigma}_y^{(j)}\right)
\end{align}
with the power-law dependence of the coupling strength, i.e., 
$J_{ij} = |i-j|^{-\alpha}$, $0<|i-j|\le L$, 
with the exponent ranging from $\alpha=0$ (fully connected all-to-all case) to $\alpha\rightarrow\infty$ (nearest-neighbors interaction).
 While the
fully connected models are now realized in experiments, the distance-dependent interactions are still a natural playground for most cases, as in the state-of-the-art quantum simulator platforms ~~\cite{Zhang2017, doi:10.1126/science.aax9743, bornet2023scalable,  Eckner2023, franke2023quantumenhanced, PhysRevX.12.011018, RevModPhys.93.025001, PhysRevLett.131.033604, PhysRevLett.129.063603}.
The above Hamiltonian is of particular importance for potential experiments aiming to generate non-classical correlations in spin chains. The behavior of the two-body Bell correlations around the critical value of the transverse-field Ising model has been studied in~\cite{Piga2019}.

\begin{figure}[t!]
  \includegraphics[width=\linewidth]{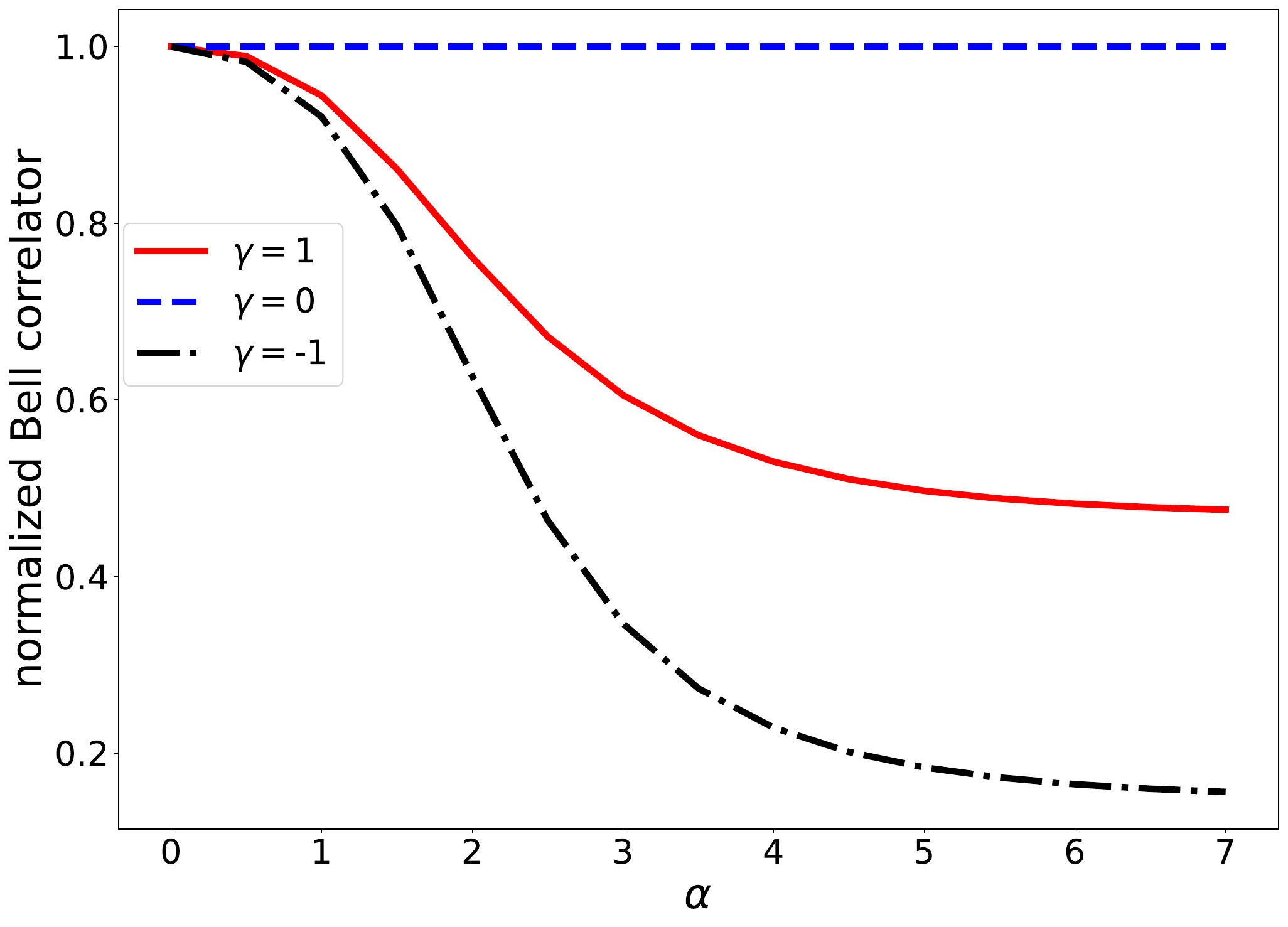}
  \caption{Many-body Bell correlator calculated with the ground state for the distance-dependent interactions with $L = 8$. 
      The dashed blue line is for $\gamma=0$, the solid red line for $\gamma=1$ and the dot-dashed black line is for $\gamma=-1$.}
    \label{fig:fig_5}
\end{figure}

Note that the Hamiltonian from Eq.~\eqref{eq:H_power_law} is absent of the permutational invariance. Hence, we must refer to the single-spin addressing Bell correlator from Eq.~\eqref{eq.elq}
or \eqref{eq.bell.q} and optimize the orientations of the $\hat\sigma_+^{(k)}$ operators independently for each $k$, which significantly increases the computational load.
In Fig.~\ref{fig:fig_5} we present  the optimized Bell correlation for the  ground state and using finite $\alpha$, normalized to the value obtained with the all-to-all case (i.e., such that can be reproduced by setting $\alpha = 0$). 
The blue dashed line shows the constant $\alpha$-independent normalized Bell correlations for $\gamma=0$, while the solid red and dot-dashed black line present a decaying Bell correlator for 
$\gamma=1$ and $-1$, respectively.

A particularly instructive case is for infinitesimally small $\gamma$'s. We start with the $\gamma = 0$ case, giving the Hamiltonian \begin{align}\label{eq.ham.J}
  \hat{H} = -\frac{1}{L}\sum_{i<j}J_{ij}\hat{\sigma}_x^{(i)}\hat{\sigma}_x^{(j)}.
\end{align}
Due to the presence of the minus sign in front, the minimal energy when the sum is maximal, i.e., when the spins {\it all'unisono} point either ``up'' or  ``down''. The doubly degenerate ground state is either
$|0\rangle_x^{\otimes L}$ or $|1\rangle_x^{\otimes L}$. 
Both are fully separable, a consequence of the classical structure of the Hamiltonian Eq.\eqref{eq.ham.J}, only involving operators $\hat{\sigma}_x$.
However, even infinitesimal $\gamma$ lifts the ground state degeneracy. As a result, the GHZ states
\begin{align}\label{eq:psi_GHZ_x}
    |\psi_\pm\rangle = \frac{1}{\sqrt{2}}(|0\rangle_x^{\otimes L} \pm |1\rangle_x^{\otimes L}),
\end{align}
become of minimal energy. Such highly entangled states give the maximal value of the correlator, see the discussion around Eq.~\eqref{eq.ghz} (for detailed discussion of the quantum Ising ground state see~\cite{Parkinson2010, mbeng2020quantum})
Note that this argument holds for all $\alpha$'s, a consequence of
the minus sign in front of the sum in the Hamiltonian~\eqref{eq.ham.J}.

The other cases, $\gamma = \pm 1$, call for a numerical-only approach and, as shown in Fig.~\ref{fig:fig_5}, the dependence of  $\alpha$ is strongly pronounced.

\subsection{Diagonal and off-diagonal disorder}

\begin{figure}[t!]
    \includegraphics[width=\linewidth]{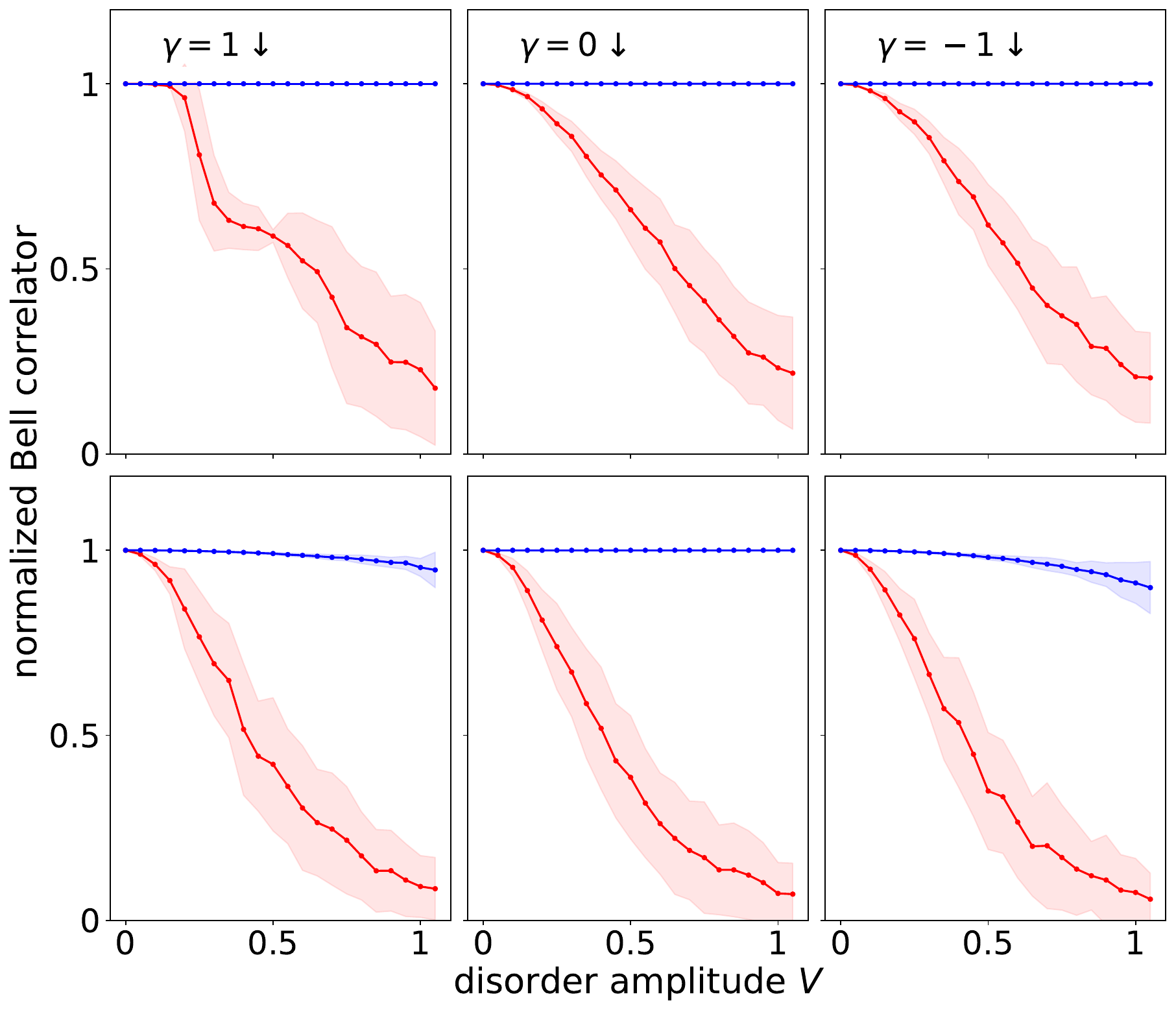}
    \caption{Relative change of the Bell correlations in the ground state, as a function of disorder amplitude. The red lines with shaded areas correspond to diagonal disorder, while the blue lines (with shaded area, not visible) correspond to off-diagonal disorder. The Bell correlations are robust to off-diagonal disorder, and preserved with diagonal disorder. Top row shows the impact of the uniform distribution ${\cal P}^{(1)}$ while the bottom row is for the experimental distribution ${\cal P}^{(2)}$. The number of spins is $L = 8$.}
    \label{fig:fig_6}
\end{figure}

In this last section, we enrich the above distance-dependent model, taking into account non-vanishing $h$, namely
\begin{align}\label{eq:H_noise}
  \hat{H} = -\frac{1}{L}\sum_{i<j}J_{ij}\left(\hat{\sigma}_x^{(i)}\hat{\sigma}_x^{(j)}+\gamma\hat{\sigma}_y^{(i)}\hat{\sigma}_y^{(j)}\right)+\sum_i h_i\hat\sigma_z^{(i)}
\end{align}
 and allowing for the presence of fluctuations of both $h_i$ 
(called the \textit{diagonal disorder}) and of the spin couplings $J_{ij}$ (the \textit{off-diagonal disorder}). Such fluctuations are inevitable in realistic configurations hence it is of importance to 
perform a numerical analysis of their influence on the Bell correlator. 

We consider two types of probability distributions that generate the noisy random variable $\epsilon$ added to either $h$ or $J_{ij}$, i.e. $h_i = \epsilon$ or $J_{ij} = J_{ji} = 1 + \epsilon$. These
are, the uniform distribution from $-V/2$ to $V/2$, i.e., 
\begin{align}
  \mathcal P^{(1)}_a(\epsilon)=\frac 1V
\end{align}
and alternatively, the probability distribution given by
\begin{align}
  \mathcal P^{(2)}_a(\epsilon)=\frac{1}{\pi\sqrt{\epsilon(V-\epsilon)}}.
\end{align}
This former type of noise was recently generated and analyzed in the experimental realization of the random-spins models in an optical cavity~\cite{Sauerwein2023}.
The Fig.~\ref{fig:fig_6} summarizes the results for the ground state and for three the cases of $\gamma$ and $\mathcal P^{(i)}_a(\epsilon)$, showing a general tendency---Bell correlations  are robust to off-diagonal disorder. 
Also, we note that Bell correlations are weakly affected by the diagonal disorder as long as its amplitude is much smaller than $V\ll 1$.

\section{Outlook and Conclusions}\label{sec:Conclusions}

This work addressed the problem of generating the strongest possible and scalable many-body quantum correlations in spin-$1/2$ chains. 
We showed that the creation of quantum states with strong many-body Bell correlations does not necessarily have to be done via dynamical protocols such as one-axis twisting. 

We showed that many-body Bell correlations are inherently present in the eigenstates of a broad family of spin-$1/2$ models described by the 
Lipkin-Meskhov-Glick model with all-to-all and power-law decaying couplings. Next, we showed that inherent many-body Bell correlations are also present in the thermal states of the LMG model, and we estimated the critical temperature above which Bell correlations vanish for the one-axis twisting Hamiltonian. Finally, we showed that the many-body Bell correlations are robust to disorder potential. We hope that this work will shed some light on the properties of the many-body Bell correlations beyond the pure state description~\cite{Fadel2018bellcorrelations, PhysRevA.105.L060201}.

The remaining question is whether there are possibilities of experimental certification of the many-body Bell correlations in the quantum systems. The measurement of the off-diagonal GHZ element of the density matrix in $L=4$ qubit system has been prepared experimentally \cite{Sackett2000}. 
Recently it was shown the many-body Bell correlations can be extracted with the quantum state tomography via the classical shadows technique~\cite{plodzien2024generation} up to $L = 10$ spins. 
Another approach is a technique based on multiple quantum coherence~\cite{garttner2017measuring,PhysRevLett.120.040402}, 
which allows to access to individual elements of the density matrix, including those that encode information about many-body Bell correlations.

\section*{Acknowledgement}
We thank Guillem M\"uller-Rigat, and Jordi Tura for reading the manuscript and their useful comments.
JCh was supported by the National Science Centre, Poland, within the QuantERA II Programme that has received funding from the European Union’s Horizon 2020 research and innovation programme under Grant Agreement No 101017733, Project No. 2021/03/Y/ST2/00195.

ICFO group acknowledges support from:
European Research Council AdG NOQIA; MCIN/AEI (PGC2018-0910.13039/501100011033, CEX2019-000910-S/10.13039/501100011033, Plan National FIDEUA PID2019-106901GB-I00, Plan National STAMEENA PID2022-139099NB, I00, project funded by MCIN/AEI/10.13039/501100011033 and by the “European Union NextGenerationEU/PRTR" (PRTR-C17.I1), FPI); 
QUANTERA MAQS PCI2019-111828-2; QUANTERA DYNAMITE PCI2022-132919, QuantERA II Programme co-funded by European Union’s Horizon 2020 program under Grant Agreement No 101017733; Ministry for Digital Transformation and of Civil Service of the Spanish Government through the QUANTUM ENIA project call - Quantum Spain project, and by the European Union through the Recovery, Transformation and Resilience Plan - NextGenerationEU within the framework of the Digital Spain 2026 Agenda; Fundació Cellex; Fundació Mir-Puig; Generalitat de Catalunya (European Social Fund FEDER and CERCA program, AGAUR Grant No. 2021 SGR 01452, QuantumCAT \ U16-011424, co-funded by ERDF Operational Program of Catalonia 2014-2020); Barcelona Supercomputing Center MareNostrum (FI-2023-3-0024);  Funded by the European Union. 

Views and opinions expressed are however those of the author(s) only and do not necessarily reflect those of the European Union, European Commission, European Climate, Infrastructure and Environment Executive Agency (CINEA), or any other granting authority.  Neither the European Union nor any granting authority can be held responsible for them (HORIZON-CL4-2022-QUANTUM-02-SGA  PASQuanS2.1, 101113690, EU Horizon 2020 FET-OPEN OPTOlogic, Grant No 899794),  EU Horizon Europe Program (This project has received funding from the European Union’s Horizon Europe research and innovation program under grant agreement No 101080086 NeQSTGrant Agreement 101080086 — NeQST);  ICFO Internal “QuantumGaudi” project;  European Union’s Horizon 2020 program under the Marie Sklodowska-Curie grant agreement No 847648; “La Caixa” Junior Leaders fellowships, La Caixa” Foundation (ID 100010434): CF/BQ/PR23/11980043. 

\bibliography{biblio}

\end{document}